# Prediction of Stable Ground-State Uranium Nitrides at Ambient and High Pressures

Dawei Zhou[†], JiaHui Yu[‡], Chunying Pu*[,†], Yuling Song[†]

[†] *College of Physics and Electronic Engineering, Nanyang Normal University, Nanyang 473061, China*

[‡] *Department of Electronics and Electrical Engineering, Nanyang Institute of Technology, Nanyang 473004, China*

**Abstract**

Uranium nitrides have been the subject of intense research owing to their potential applications as advanced nuclear fuels. However, the phase diagram of the U-N system at low temperature and high pressure still remains unclear. In this paper, we explore extensively the phase diagram of the U-N system up to 150 GPa based on first-principles swarm structure searches. The phase diagrams of the experimentally known stoichiometries like $U_2N_3$ and $UN_2$ are refined. At zero temperature and pressure, the experimentally observed *CaF*$_2$-type $UN_2$ is found to transform into another new *I*4$_1$/*amd*-type $UN_2$, which is related to the dynamical instability originated from *Peierls* mechanism. Two new stable high-pressure phases of $U_2N_3$ and $UN_2$ are identified for the first time. Besides, several new chemical stoichiometries ($UN_4$, $UN_3$, $U_3N_5$ and $U_2N$) are found to have stability fields on the U-N phase diagram. The pressure-induced phase transitions for the U-N system are further investigated. The peculiar structural features such as $N_2$-dimers, planar SO3-like N(N)$_3$ units, non-coplarnar zigzag $N_4$ units, and zigzag U chains are found in U-N compounds under pressure. Our results on the structure exploring provide a better understanding of the structural characteristics and physical properties of uranium nitrides under pressure.

# 1. Introduction

Actinide nitrides continue to attract intensive attention due to their special technological and scientific importance in recent years. Firstly, these compounds have been considered as the primary candidate fuel for the Generation-IV fast nuclear reactors.[1-2] Compared to standard oxide fuels, nitride fuels have several advantages, such as high melting point, high metal density, high thermal conductivity and good compatibility with the sodium coolant.[3-7] The above-mentioned excellent characteristics of nitride fuels allow reactor operation at elevated temperatures, thus improving effectively the energy conversion efficiency. Secondly, due to the unique behavior of 5f electrons, actinide based materials usually exhibit special physical and chemical properties. Some of the 5f electrons in actinide based materials are found to be neither fully localized nor completely itinerant, which are believed to have dual (itinerant and localized) natures. Furthermore, strong Coulomb correlation is found among the partially filled 5f electrons. The unique nature of the 5f-electrons is believed to be responsible for many interesting properties like the heavy fermion state, unconventional superconductivity, multipole orderings, etc. observed in some uranium compounds.[8-12] Among those Actinide nitrides, uranium nitrides are typical example of them, and numerous experimental and theoretical works have focused on them. The physical and chemical properties of uranium nitrides such as the reactivity features, electronic structure, lattice dynamics, surface properties, structural phase transition, magnetic, elastic and thermodynamic properties have been investigated in the previous work.[13-33]

However, the phase diagram of the uranium nitrides at low temperature and high pressure remains unknown. Up to now, there are mainly four experimentally reported ordered uranium nitride compounds,[26, 27] eg. UN, $\alpha$-$U_2N_3$, $\beta$-$U_2N_3$, and $UN_2$, the crystal structures of which are NaCl-type, $Mn_2O_3$-type, $La_2O_3$-type, and $CaF_2$-type, respectively. According to the previous work[28], $UN_2$ can be formed at 500 ℃, and then decomposes into $\alpha$-$U_2N_3$ at 675℃, while $\alpha$-$U_2N_3$ begins to decompose into UN at 975℃. We noticed that the nitrogen content in the experimentally reported U-N compounds increases as temperature decreases, so temperature plays a very important role on synthesizing U-N compounds, and there may exist some other stable stoichiometries of the U-N system at low temperature. Furthermore, among the already known U-N compounds, only the pressure-induced phase transition of UN has been investigated[17,22,29], while the high-pressure phases of $U_2N_3$ and $UN_2$ still remain unknown. Even at zero pressure, the experimentally observed *CaF*$_2$-type $UN_2$ is reported to be unstable at zero temperature and a negative frequency is found in its phonon curves in a recent theoretical work.[30] Besides, considering the complex chemical bonds of uranium and nitride atoms, we also believe that there may exist some other stable stoichiometric of the U-N system. So as the one of the most important compounds as we have mentioned above, it is necessary to refine phase diagram of the U-N compounds at low temperature and high pressure.

In this paper, we explored the possible stable stoichiometries of the U-N systems with $U_mN_n$ ($UN_4$, $UN_3$, $UN_2$, $U_3N_5$, $U_2N_3$, UN, and $U_2N$) compositions from 0 to 150 GPa using a swarm intelligence-based structural prediction based on the first-principles calculations. As we have expected, pressure leads to the structural diversity of the U-N system, and many fascinating chemical bonding characteristics were found in U-N system. A new stable zero-pressure phase of $UN_2$ and several stable high-pressure phases of $U_2N_3$ and $UN_2$ were identified for the first time. Many new stable stoichiometries ($UN_4$, $UN_3$, $U_3N_5$ and $U_2N$ ) were found to have stability fields on the U-N phase diagram. The structural stability, chemical bonding characteristics, electron structure, phase transition mechanism and the morphology of nitrogen atoms in those newly found structures were investigated systematically. Since uranium nitrides have been purposed to be used in future reactors and subcritical systems, our results on the structure exploring are helpful to design new typical nitride fuels, and also helpful to better understand the structural and electronic characteristics of actinide compounds under pressure.

## 2. Computational Methods

Our structural prediction approach is based on a global minimization of free energy surfaces merging *ab initio* total-energy calculations with CALYPSO (Crystal structure AnaLYsis by Particle Swarm Optimization) methodology as implemented in the CALYPSO code.[34-36] The crystal structures of various stoichiometry $U_mN_n$ ($UN_4$, $UN_3$, $UN_2$, $U_3N_5$, $U_2N_3$, UN, and $U_2N$) were searched with simulation cell sizes of 1-4 formula units (f.u.) at 0, 50, 100, and 150 GPa, respectively. In the first step, random structures with certain symmetry are constructed in which atomic coordinates are generated by the crystallographic symmetry operations. The first generation of structures created randomly. At each step, only 60% of each generation was used to produce the next-generation structures by PSO (Particle Swarm Optimization). 40% of the structures in the new generation are randomly generated. In most cases, structural searching simulations for each calculation were stopped after generating ~1000 structures (e.g., about 20~30 generations). Due to the onset of 5$f$-electron localization phenomena, the normal LDA or GGA usually underestimates the strong on-site Coulomb repulsion of the 5$f$–electron. Fortunately, we can use the LDA/GGA+$U$ method developed by Dudarev *et al.*[37] to effectively remedy the failures originated from normal LDA/GGA. However, choosing an appropriate Hubbard $U_{eff}$ value is another problem. According to previous work [18, 21, 31-33], we found that for the experimentally known U-N system such as UN, $UN_{1.5}$, and $UN_2$, the physical properties of them can be described well by choosing the Hubbard U parameter around 2 eV, including the electronic structures, bonding properties, the phonon dispersions, equation of state and pressure-induced phase transition. In fact, a different U value may slightly modify the transition pressure, but generally does not change the pressure-induced structural sequence, so for all calculations in this paper, the GGA+U approach and an effective Coulomb interaction $U_{eff}$=2 eV, for unranium 5$f$ orbitals, were used.

Once a new structure was obtained from CALYPSO, the structural relaxation and properties calculations were implemented in the VASP code,[38, 39] which were carried out using the all-electron projector-augmented wave (PAW) [40] method within the framework of density functional theory. The exchange correlation potential were treated within the generalized gradient approximation (GGA) of Perdew-Burke-Ernzerhof (PBE). [41] The U and N pseudopotentials were generated by taking $6s^2 6p^6 5f^3 6d^1 7s^2$ and $2s^2 2p^3$ as the valence electrons configurations. The plane-wave kinetic energy cutoff of 500 eV for the expansion of the wave function into plane waves was chosen to ensure that the results are in accordance with the convergence criteria. The Monkhorst-Pack $k$-mesh with grid spacing of $2\pi \times 0.03$ Å$^{-1}$ was chosen to ensure that the total energy calculations are well-converged to better than 1 meV/atom.

To ensure that the obtained structures are dynamically stable, we calculated the phonon frequencies throughout the Brillouin zone using the finited-displacement implemented in the Phonopy code. [42] To elucidate the bonding information in these new phases, we adopted a variant of the familiar COHP approach that stems from a PW calculation and was dubbed "projected COHP" (pCOHP). [43] In this approach, all the projection and analytic methods have been implemented in a standalone computer program, LOBSTER, which can give us access to the projected COHP curves and we can used to visualize the chemical-bonding information in our DFT+$U$ calculations. We also calculated the charge density difference for the predicted U-N systems in this paper, which is defined as: $\Delta\rho = \rho_{sc} - \rho_{atom}$, where $\rho_{sc}$ is the total charge density obtained after self-consistent calculations. $\rho_{atom}$ is the total charge density obtained after non-self-consistent calculations.

## 3. Results and discussion

### 3.1 Phase stabilities and structures of U-N compounds

We have performed extensive structural searches on the U-N system for various stoichiometry $U_m N_n$ ($UN_4$, $UN_3$, $UN_2$, $U_3N_5$, $U_2N_3$, UN, and $U_2N$) compositions at 0K and selected pressures of 0, 50, 100, and 150 GPa. The stability of $U_m N_n$ can be quantified by constructing the thermodynamics convex hull at the given pressure, which is defined as the formation enthalpy per atom of the most stable phases at each

stoichiometry:

$$\Delta H_f(U_m N_n) = [H(U_m N_n) - mH(U) - nH(N_2/2)]/(m+n) \tag{1}$$

where $\Delta H_f$ the enthalpy of formation per atom and $H$ is the calculated enthalpy per stoichiometric unit for each compounds. The enthalpies $H$ for $U_m N_n$ for the most stable structures are searched by the CALYPSO method at the desired pressures. The known structures of $Pa$-3, $P4_12_12_1$, and $I2_13$ for N, [44] and $Cmcm$ for U [45] in their corresponding stable pressure were correctly predicted. Any structure with its enthalpy lying on the convex hull is considered to be the thermodynamically stable.

The convex hulls at selected pressure are depicted in Fig.1(a). The lattice parameters of all of the predicted structures are listed in the Supporting information (Table S1). From the calculated convex hull surface at zero pressure, we can see that there are four stable thermodynamically structures, i.e., NaCl-UN, $\alpha$-$U_2N_3$, $Pbam$-$U_3N_5$ and $I4_1/amd$-$UN_2$. Both NaCl-UN and $\alpha$-$U_2N_3$ are experimentally known structures. We noticed that $\beta$-$U_2N_3$ has a higher energy than $\alpha$-$U_2N_3$, so it does not lie on the convex curves at zero pressure. For $UN_2$, a new tetragonal $I4_1/acd$ phase is found to be energetically stable than the experimentally observed $CaF_2$ phases. Furthermore, it is very significant, and in fact surprising, that a new stoichiometry $U_3N_5$ is found to be stable at ambient pressure.

At elevated pressures, the U-N system further shows its unexpected structural diversity. It is exciting to note that N-rich compositions $UN_4$, $UN_3$ and U-rich composition $U_2N$ are stable thermodynamically above 50 GPa. As pressure increases above 150 GPa, $UN_2$ is slightly above the convex hull and may become a metastable phase. For $U_2N_3$ and $U_3N_5$, they show complex behaviors under pressure and may undergo metastable processes due to the slight deviations from convex hull at 50 GPa and 150 GPa. Except for $UN_3$, all the other U-N compounds undergo a series of structural phase transformations under pressure. The stable pressure ranges of various phases for the stable $U_m N_n$ compounds are depicted in Figure 1(b). Especially, we showed that UN undergoes a phase transition from $Fm$-$3m$ to $R$-$3m$ phase at a pressure of about 31 GPa, in agreement with the experimental measurement [17] and other calculated results. [22, 29].

To investigate the structural characteristics of the newly found $U_mN_n$ compounds, all the predicted structures are shown in Figure 2, and the charge density difference maps (CDDM) of the characteristic N-rich compounds ($UN_4$ and $UN_3$) and U-rich compound $U_2N$ are given in Figure 3, while that of the other predicted compounds ($UN_2$, $U_3N_5$, and $U_2N_3$) are given in Fig.S1.

For the N-rich compounds $UN_4$, three stable phases are found, i.e., $P2_1/C$, $P2_1$ and $P6_3/mmc$. As pressure increases, $P2_1/C$ transforms into $P2_1$ at about 51.2 GPa, and then into $P6_3/mmc$ at 135.0 GPa. The crystal structure of $P2_1/C$ (Fig.2a) contains diatomic $N_2$ dimers, and the calculated bond length between the N-N bonds is 1.274 Å. While for $P2_1$ (Fig.2b) structure, it is interesting to note that the nitrogen atoms form non-coplanar zigzag-shaped $N_4$ chains. The bonding characteristics of the $N_4$ chains can be seen clearly from CDDM of $P2_1$ phase (Fig.3b). Although all kinds of nitrogen chains or clusters have been reported in nitrogen-rich compounds, as far as we known, this kind of non-coplanar zigzag-shaped $N_4$ chain is firstly found. The $P6_3/mmc$ $UN_4$ at 150 GPa is displayed in Fig.2c, the nitrogen atoms in which form a planar SO3-like $N(N)_3$ units. Each planar $N(N)_3$ moiety possesses $D_{3h}$ symmetry with N-N bond lengths of 1.307 Å. The detailed $N(N)_3$ units are illustrated by the CDDM of $P6_3/mmc$ phase as shown in Fig.3c. We noticed that the similar $N(N)_3$ unit was also reported in the high-pressure phases of the Mg-N compounds.[46]

$UN_3$ crystalizes in a tetragonal structure. Similar with P2$_1$/c $UN_4$, the $N_2$ dimers are also discovered in this structure. The N-N distance of $N_2$ dimers is 1.369 Å, and the two $N_2$ are separated by a large distance (2.249Å) forming a rectangular shape. The N-N distances in these N-rich compounds exhibit the covalent bond character and produce strong interatomic interactions that can be seen from CDDM (Figure 3a-d), playing a key role in stabilizing the N-rich structure.

The crystal structure of $UN_2$ is reported experimentally as $CaF_2$-type structure at zero pressure, however, K.O.Obodo etal[30] found that there are small imaginary phonon frequencies in the phonon curves of the ground-state $CaF_2$ phase at zero pressure. So in theory, $CaF_2$-type $UN_2$ is not a stable structure at zero temperature and pressure. Using the structure predictions, we found a new stable tetragonal $UN_2$ (space group $I4_1/acd$, 4 formula units per cell, Figure 2e) at zero pressure. The newly found $I4_1/acd$ phase is energetically stable than the $CaF_2$ phase (lower 2.6 meV/atom), which is also dynamically stable as we will discuss latter. Interestingly, as pressure increases, the $I4_1/acd$ phase transforms to $CaF_2$ phase at about 45.0 GPa, and then to a orthorhombic structure (space group *Pnma*, Figure 2f) at pressure 70.6 GPa. We noticed that the all nitrogen atoms in the $UN_2$ have a similar morphology, which exist as isolated atoms.

$U_3N_5$ is a new stoichiometry of the U-N compounds which can exist stably at zero pressure. It adopts an orthorhombic structures with space group of *Imm*2 (Figure 2g), and then transforms to a hexagonal structure with space group *R-3m phase*(Figure 2h) at 8.2 GPa. The *R-3m* phase may exist as a metastable structure at 50.0 GPa due to a slight enthalpy deviation from convex hull. However, as pressure increases to 100 GPa, a newly found orthorhombic *Pbam* structure (Figure 2i) resides on the convex hull again, implying the thermodynamic stability of $U_3N_5$. For *Imm*2-$U_3N_5$, each nitrogen atom is bonded by four uranium atoms forming N-centred regular tetrahedron. For *R-3m*-$U_3N_5$, there are two types of N atoms occupying the 3*b* and 6*c* Wyckoff sites, respectively. Each 3*b* N atom is coordinated by six U atoms, while the 6*c* N atom coordinated by four U atoms. For *Pbam*-$U_3N_5$, three types of N atoms occupy 4*g*, 4*h*, and 2*a* Wyckoff sites, respectively. Each 4*g* N atom is surrounded by four U atoms forming pyramid units, and 4*h* N atom is bonded with four U atoms forming asymmetric tetrahedrons, while 2*a* N atoms is coordinated by four U atoms forming planar N-centered $U_4$ units.

The $U_2N_3$ compounds have two phases found experimentally, i.e., α-$U_2N_3$ and β-$U_2N_3$. It is found that α-$U_2N_3$ has a lower energy than β-$U_2N_3$ at zero pressure. As pressure increases, α-$U_2N_3$ is found to transform to β-$U_2N_3$ at 5.2 GPa. Upon further compression, a monoclinic structure of $U_2N_3$ is predicted to be stable at about 73.4 GPa as shown in Figure 2g (space group *C*2/*m*, 2 formula units per cell). Similar with $UN_2$, the isolated nitrogen atoms are also found in $U_2N_3$.

For U-rich compounds, $U_2N$ becomes a stable stoichiometry of U-N system above 50 GPa, and it first adopts a orthorhombic structure with space group of *Cmca* (Figure 2k) and then undergoes a pressure-induced phase transition to a monoclinic phase with space group of *C*2/*m* (Figure 2l) at 98.0 GPa. Interestingly, both structures consist of one dimensional zigzag U chains, which play an important role in stabilizing the U-rich compounds. Clearly covalent bonds are found among the U atoms (See the CDDMs in Fig.3(e) and Fig.3(f), respectively), leading to the formation of the zigzag U chains.

The values of relative volume change associated with compressions for all U-N compounds are obtained and plotted in Fig.4. Except for the phase transition from $I4_1/amd$ to $CaF_2$ in $UN_2$, it can be seen clearly that the volumes of all the other U-N compounds change discontinuously during the transitions, indicating that the phase transitions have a first order nature. Especially, we noticed that $U_2N$ undergoes a pressure-induced phase transition with a large volume collaps 38.7%. On the contrary, when the phase transition from $I4_1/amd$ to $CaF_2$ in $UN_2$ occurs, the volume remains almost unchanged. In fact, the transition from $I4_1/amd$ to $CaF_2$ $UN_2$ is related to a soft mode transition, which has a second-order nature as we will discuss later. The above calculated *P-V* curves provide fundamental date which can be easy to detect in further experiments.

**3.2 Structural dynamical stability**

Except for the thermodynamic stability, the dynamical stability is also very important. If a crystal does not meet the dynamical stability criterion, it cannot exist stably in theoretically. Thus we calculated the phonon dispersions in the whole Brillouin zone for all the newly found structures of the U-N system at selected pressure, which are shown in Fig.5. The nonexistence of imaginary frequencies indicates the dynamic stabilities of all the new predicted structures.

Now we further to investigate the phase transition mechanism from $I4_1/amd$ to experimentally observed $CaF_2$ in $UN_2$, which is found to be related to the dynamical instability. We calculated the phonon curves of $CaF_2$ structure using a conventional cell at zero pressure. In agreement with the previous work[30], an imaginary frequency at X-point (0.5, 0.0, 0.0) is found as shown in Figure 6a. As pressure increases to 50 GPa, the imaginary frequency at X-point disappears, so $CaF_2$ phase becomes a stable structure, consistent with our enthalpy curve results. In fact, we found that the crystal structures of $I4_1/amd$ and $CaF_2$ $UN_2$ are very similar, and if the nitrogen atoms in $I4_1/amd$ phase does not distort and range in a line along *x* direction, then the $I4_1/amd$ phase will become the $CaF_2$ phase. Theoretically, the appearance of soft phonon modes usually leads to a distortion of the crystal, so we investigated the unstable mode at X-point in detail. It is found that the unstable mode at X-point is just comprised of predominantly the movement of N atoms along $x$ direction. The potential energy curve for this mode is calculated by giving atoms displacement according to this mode with increasing amplitude, the potential energy curve for this mode is shown in Fig.6 b. At zero pressure, a potential well is found, so the atoms displaces along this mode will result in a new stable structure. Surprisingly, the new distorted structure is just the $I4_1/amd$ structure. The above phase transition is just so-called "soft mode phase transtion". Moreover, as pressure increases, the potential well disappears at 50 GPa and thus $CaF_2$ becomes stable. So we can conclude that the phase transition from $CaF_2$ to $I4_1/amd$ at zero temperature and pressure is related to the dynamical instability, which leads to the distortions of nitrogen atoms along $x$ direction. The distortions of atoms in a crystal generally increase the coulomb repulsion energy among the atoms, but if the distortions can change the crystal electronic structure and the one-electron energy sum is reduced enough, then the distorted structure may exist stably. The above-mentioned phase transition mechanism is so-called Peierls mechanism[47]. There are some distorted structures found in actinide compounds such as $PuO_2$, $U_2Ti$, and Peierls mechanism is found to be responsible for the distortions.[48-49] To further reveal the phase transformation mechanism, we calculated the electron density of states for both $CaF_2$ and $I4_1/amd$ structures, which are shown in Figure.6c. It is found that Fermi energy decreases from 9.4587 eV for $CaF_2$ phase to 9.4183 eV for the distorted $I4_1/amd$ phase. The shift of bands to lower energy for the distorted $I4_1/amd$ phase is also found. So we believe the physically driven force for the phase transition is also related to the Peierls mechanism. We further simulated the X-ray diffractometry (XRD) pattern of both two $UN_2$ structures with a wavelength of 1.54056 Å at 0 GPa, which are shown in Fig.6 d. Different from $CaF_2$ phase, there are additional five small peaks as circled in $I4_1/amd$ phase, which can provide useful information for future experiments to identify the crystal structure.

**3.3 Electronic structures and chemical bonds**

To get an in-depth understanding of the role of the 5$f$-electrons on $U_mN_n$ compounds, we have carried out a detailed analysis of electronic structure. In Fig.7 we calculated the band structures, total and projected density of states (DOSs) of the new $U_mN_n$ compounds with *GGA+U* methods at selected pressure. It can be seen that except for *I4$_1$/acd* UN$_2$ at 0 GP (Fig 7e), all of the other structures exhibit metallic behavior. From the electronic total DOSs and projected DOSs, we can see that the metallic character with the Fermi level is mainly dominated by the 5*f* states with some degree of hybridization with N-2*p* and very minor U-6*d* orbital contributions. The appearance of U 5*f* states near the Fermi level in studied $U_mN_n$ compounds indicates the U 5*f* states preserve the band character and further confirm the itinerant character.

As insulating character is observed for UN$_2$ at 0 GPa in the *I4$_1$/acd* phase (Fig 7e). The calculated band gap is 0.80 eV for UN$_2$ at 0 GPa in *I4$_1$/acd* phase, which is comparable to the band gap of 0.94 eV obtained in the *CaF$_2$* phase by Weck et al.[21]. In the semiconductor *I4$_1$/acd* phase, the valence band near the Fermi level is dominated by N-2*p* contributions, while U-6*d* and U-5*f* characters play only small roles. Some degree of hybridization of the N-2*p* and U-5*f* appear at the top of the valence band. The bottom of the conduction band is still strongly marked by U -5*f* orbitals, while the N-2*p* orbital contribution tends to vanish.

To understand the chemical bonding of these compounds, we performed crystal orbital Hamilton population (COHP) for all U-N system. COPH is a bond-detecting tool for solids and molecules, and can provide a straightforward view into the orbital-pair interactions. COHP partitions the band structure energy (in term of the orbital pair contributions) into bonding, non-bonding and anti-bonding contributions for a given pair-wise interaction. Conventionally, the negative COHP is plotted as a function of energy where the positive part represents bonding interaction and the negative part the anti-bonding interactions. Fig.S2 presents the CHOP for U-N, U-U, and N-N interactions in all newly predicted $U_mN_n$ compounds at selected pressures, respectively. Below Fermi level, the bonding interaction is originates from U-N and N-N interactions in N-rich compounds UN$_4$ and UN$_3$, where the bonding interaction is derives from U-N and U-U interactions in U-rich compound U$_2$N. In other predicted structures of stoichiometry compounds, the main interaction originated from U-N interactions.

To further analyze the interactions between U-N in details, Fig.8 presents the total and partial COHPs of U-N for all newly predicted $U_mN_n$ phases. From total COHPs, it is seen that except for $P2_1/C$ phase of $UN_4$, all U-N show only anti-bonding interactions above the Fermi level. Below the Fermi level, both bonding and anti-bonding interactions are present for partial U-N interactions. Especially, the bonding interactions originates from U-5$f$ –N 2$p$ and U 6$d$-N 2$p$ hybridizations, while the anti-bonding interactions are contributed by U 6$s$–N 2$p$ and U 6$p$-N 2$p$ hybridizations.

## 4. Conclusions

Unbiased structure searching and density functional total energy calculations were performed to explore phase stabilities and crystal structures of the U-N compounds up to 150 GPa. We found that U is able to react with N at experimentally accessible pressures by forming various stoichiomitric U-N compounds including semiconductor and metals. The phase diagrams of the experimentally observed stoichiometries $U_2N_3$ and $UN_2$ at low temperature and high pressure were refined. For $U_2N_3$, $α$-$U_2N_3$ is found to transform to $β$-$U_2N_3$ at 5.2 GPa, and then into a newly found monoclinic structure at about 73.4 GPa. For $UN_2$, the dynamical instability of the experimental observed $CaF_2$ phase at zero temperature and pressure results in the distortions of nitrogen atoms, leading to a phase transition to a newly found $I4_1/amd$ phase. Furthermore, the phase transition to $I4_1/amd$ phase was revealed to be related the Peierls mechanism. Upon further compression, the $I4_1/acd$ phase transforms to the $CaF_2$ phase at 45.0 GPa, and then to a orthorhombic structure at pressure 70.6 GPa. Besides, a new compound $U_3N_5$ is found to can be formed at zero pressure, and more new compounds $UN_4$, $UN_3$ and $U_2N$ were discovered to be thermodynamically stable at certain pressure ranges. The structural characteristics and the morphology of nitrogen atoms in the U-N system were investigated in detail. For UN, $UN_2$, $U_3N_5$ and $U_2N_3$，the nitrogen atoms of them mainly exist as isolated atoms. While for the N-rich compounds $UN_4$ and $UN_3$, the nitrogen atoms exhibit more exotic forms. As far as we know, the diatomic $N_2$ and $N(N)_3$ units were found for the first time in U-N compounds. More interestingly, a unique non-planar zigzag-shaped N$_4$ chain in the *P*2$_1$ phase of UN$_4$ was firstly identified. In the U-rich compounds U$_2$N, U atoms are found to form special zigzag chains in covalent bond forms in the *Fmmm* and *C*2*/m* phases. COHP function of U-N interactions in U-N compounds reveals that the bonding interactions below the Fermi level originates from U 5*f* -N 2*p* and U 6*d*-N 2*p* hybridizations, while the anti-bonding interactions is derived from U 6*s*-N 2*p* and U 6*p*-N 2*p* hybridizations. The current theoretical predictions showed the structural diversity and new structural characteristics of uranium nitrides under pressure, and will most likely promote further experimental and theoretical investigations on the U-N nuclear fuels.

## Acknowledgement

This work is supported in China by the National Natural Science Foundation of China (Grant Nos. 51501093, 41773057, U1304612, and U1404608), Science Technology Innovation Talents in Universities of Henan Province (No.16HASTIT047), Young Core Instructor Foundation of Henan Province (No. 2015GGJS-122).

## Supporting Information

Detailed descriptions of the structural parameters, charge difference densities, COHPs for N-N, U-N, and U-U of the newly predicted U-N compounds at selected pressure.

# Figure Captions

**Fig.1**. (Color online) (a) Predicted formation enthalpies of $U_mN_n$ compounds with respect to elemental U and N under pressure. (b) Schematic representation of phase diagram for $U_mN_n$ compounds.

**Fig.2**. Newly predicted structures of $U_mN_n$ at selected pressures corresponding to where they are thermodynamically stable, or in the case of $R\text{-}3m\text{-}U_3N_5$ metastable at 50 GPa. The lattice parameters of all of the structures are listed in the Supporting information (Table S1) (a) $UN_4$ at 50 GPa in the $P2_1/C$. (b) $UN_4$ at 100 GPa in the $P2_1$. (c) $UN_4$ at 150 GPa in the $P6_3/mmc$. (d) $UN_3$ at 100 GPa in the $P4_2/mnm$. (e) $UN_2$ at 0 GPa in the $I4_1/acd$. (f) $UN_2$ at 100 GPa in the $Pnma$. (g) $U_3N_5$ at 0 GPa in the $Imm2$. (h) $U_3N_5$ at 50 GPa in the $R\text{-}3m$. (i) $U_3N_5$ at 100 GPa in the $Pbam$. (j) $U_2N_3$ at 150 GPa in the $C2/m$. (k) $U_2N$ at 50 GPa in the $Cmca$. (l) $U_2N$ at 100 GPa in the $C2/m$. In all structures, the small blue and large gray balls represent N and U atoms, respectively.

**Fig.3** Difference charge density maps of N-rich ($UN_4$ and $UN_3$) and U-rich ($U_2N$) compounds. (a) $UN_4$ at 50 GPa in the $P2_1/C$. (b) $UN_4$ at 100 GPa in the $P2_1$. (c) $UN_4$ at 150 GPa in the $P6_3/mmc$. (d) $UN_3$ at 100 GPa in the $P4_2/mnm$. (e) $U_2N$ at 50 GPa in the $Cmca$. (f) $U_2N$ at 100 GPa in the $C2/m$.

**Fig.4** Relative unit cell volume $V/V_0$ in newly predicted structures of $U_mN_n$ compounds including UN compounds as a function of external pressures.

**Fig.5** Phonon dispersion of newly predicted structures of $U_mN_n$ compounds at selected pressures.

**Fig.6** (a) The phonon dispersion of UN2 at 0GPa, 25 GPa, and 50 GPa; (b) (c) density of states of $UN_2$ in $I4_1/acd$ and $Fm\text{-}3m$ phases. The corresponding Fermi energy in both cases is represented by dashed line. (d) The simulated XRD for $UN_2$ in $I4_1/acd$ and $Fm\text{-}3m$ phases. (a) The simulated XRD patterns of distorted-3R and distorted-1T $ReS_2$ with a wavelength of 1.54056 Å (which may have a better performance than $\lambda$

=0.4959Å in distinguishing these two phases) at 10 GPa

**Fig.7.** The band structures and the density of states (DOSs) in newly found $U_mN_n$ compounds at selected pressures.

**Fig.8**. COHP functions for U-N interactions in newly-predicted UN compounds at selected pressure. (a) $UN_4$ at 50 GPa in the *P2_1/C*. (b) $UN_4$ at 100 GPa in the *P2_1*. (c) $UN_4$ at 150 GPa in the *P6_3/mmc*. (d) $UN_3$ at 100 GPa in the *P4_2/mnm*. (e) $UN_2$ at 0 GPa in the *I4_1/acd*. (f) $UN_2$ at 100 GPa in the *Pnma*. (g) $U_3N_5$ at 0 GPa in the *Imm2*. (h) $U_3N_5$ at 50 GPa in the *R-3m*. (i) $U_3N_5$ at 100 GPa in the *Pbam*. (j) $U_2N_3$ at 150 GPa in the *C2/m*. (k) $U_2N$ at 50 GPa in the *Cmca*. (l) $U_2N$ at 100 GPa in the *C2/m*.

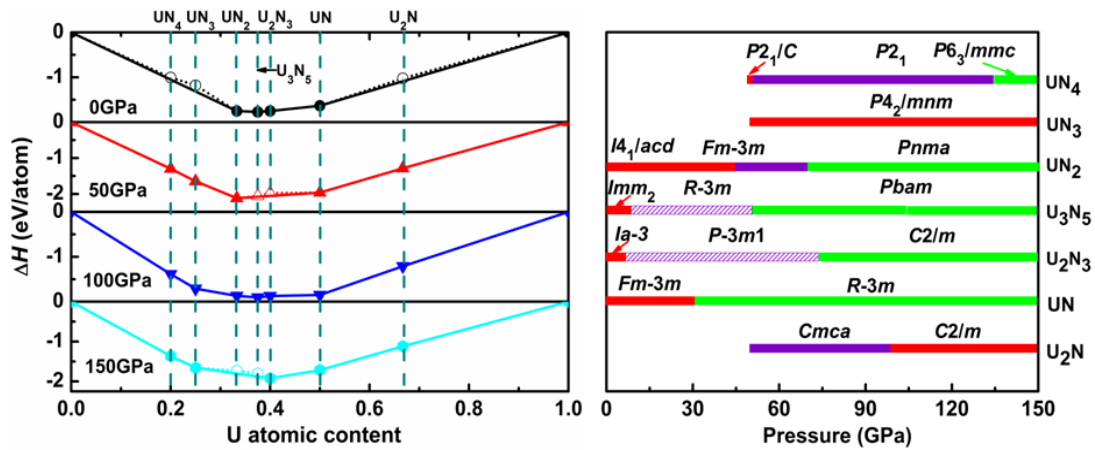

Fig1. (Color online) (a) Predicted formation enthalpies of $U_mN_n$ compounds with respect to elemental U and N under pressure. (b) Schematic representation of phase diagram for $U_mN_n$ compounds.

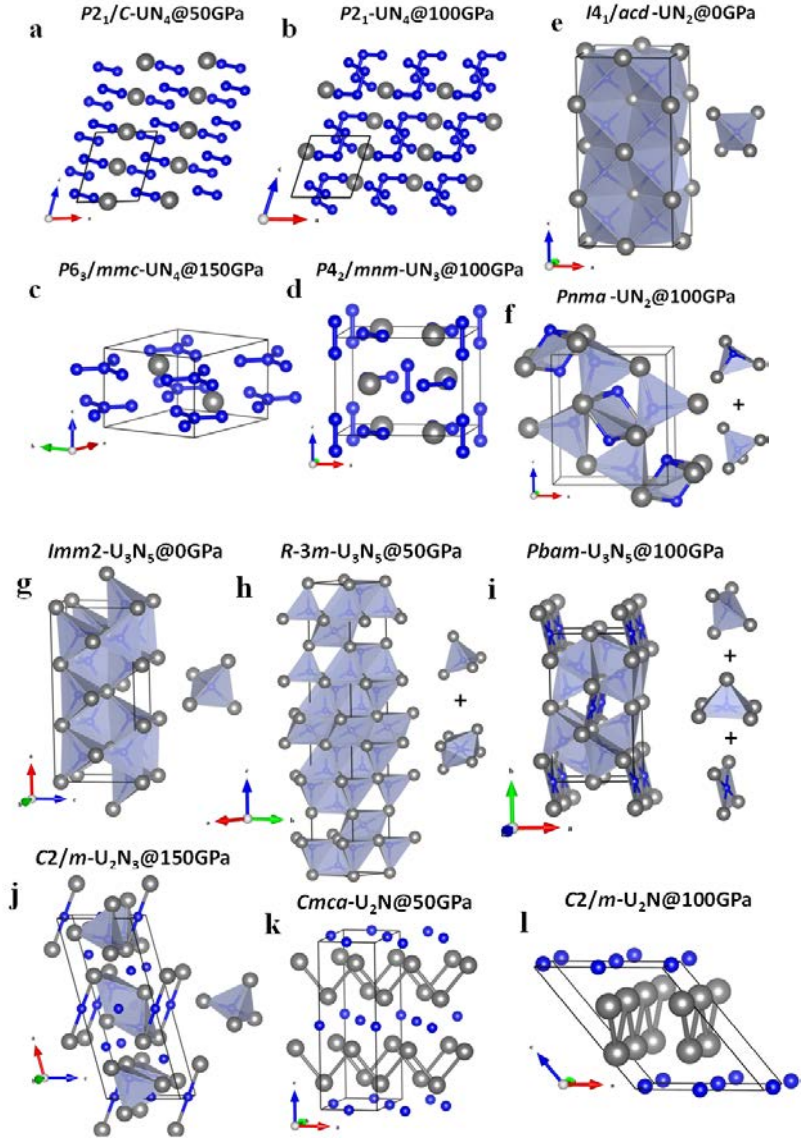

Fig.2. Newly predicted structures of $U_mN_n$ at selected pressures corresponding to where they are thermodynamically stable, or in the case of *R-3m*-$U_3N_5$ metastable at 50 GPa. The lattice parameters of all of the structures are listed in the Supporting information (Table S1) (a) $UN_4$ at 50 GPa in the $P2_1/C$. (b) $UN_4$ at 100 GPa in the $P2_1$. (c) $UN_4$ at 150 GPa in the $P6_3/mmc$. (d) $UN_3$ at 100 GPa in the $P4_2/mnm$. (e) $UN_2$ at 0 GPa in the $I4_1/acd$. (f) $UN_2$ at 100 GPa in the *Pnm*a. (g) $U_3N_5$ at 0 GPa in the *Imm2*. (h) $U_3N_5$ at 50 GPa in the *R-3m*. (i) $U_3N_5$ at 100 GPa in the *Pbam*. (j) $U_2N_3$ at 150 GPa in the *C2/m*. (k) $U_2N$ at 50 GPa in the *Cmca*. (l) $U_2N$ at 100 GPa in the *C2/m*. In all structures, the small blue and large gray balls represent N and U atoms, respectively.

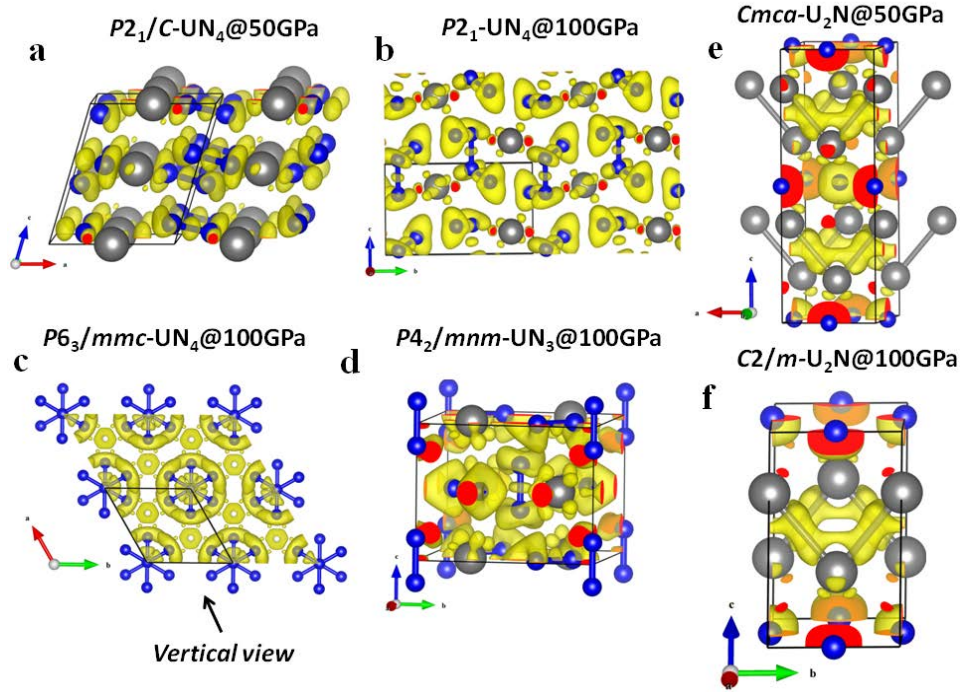

Fig.3 Difference charge density maps of N-rich (UN$_4$ and UN$_3$) and U-rich (U$_2$N) compunds. (a) UN$_4$ at 50 GPa in the *P*2$_1$/*C*. (b) UN$_4$ at 100 GPa in the *P*2$_1$. (c) UN$_4$ at 150 GPa in the *P*6$_3$/*mmc*. (d) UN$_3$ at 100 GPa in the *P*4$_2$/*mnm*. (e) U$_2$N at 50 GPa in the *Cmca*. (f) U$_2$N at 100 GPa in the *C*2/*m*.

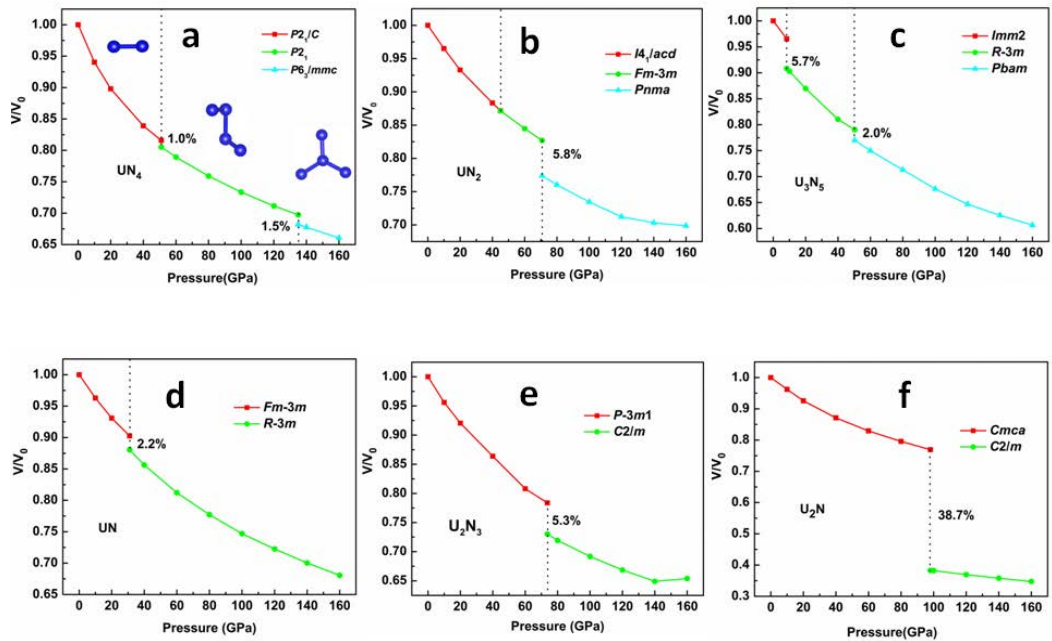

Fig.4 Relative unit cell volume V/V$_0$ in newly predicted structures of U$_m$N$_n$ compounds including UN compounds as a function of external pressures.

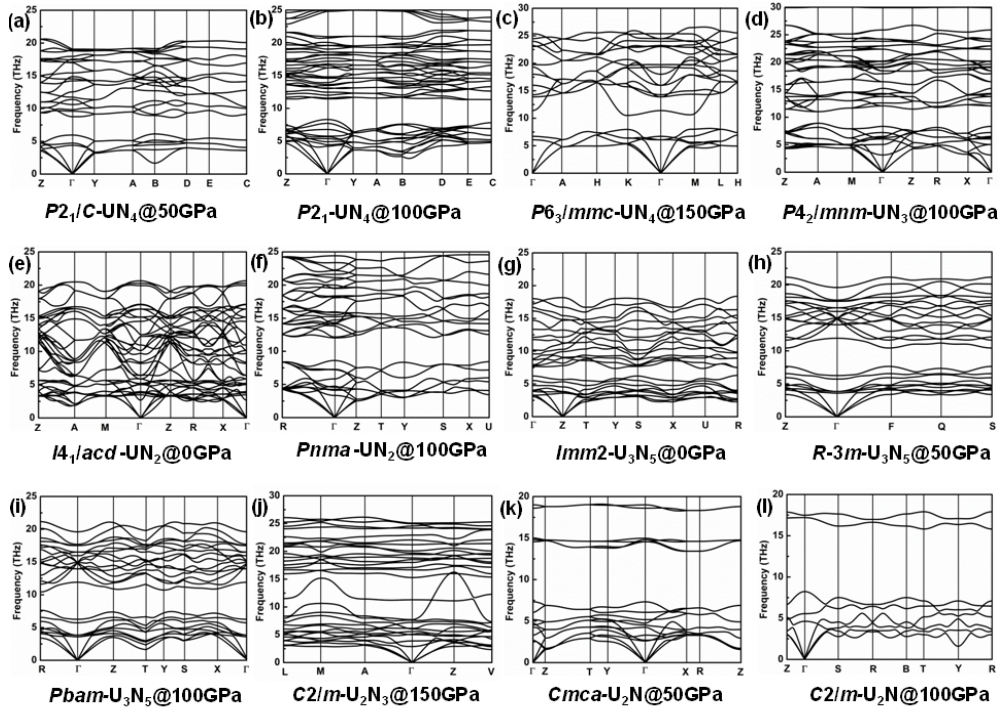

Fig.5 Phonon dispersion of newly predicted structures of U$_m$N$_n$ compounds at selected pressures.

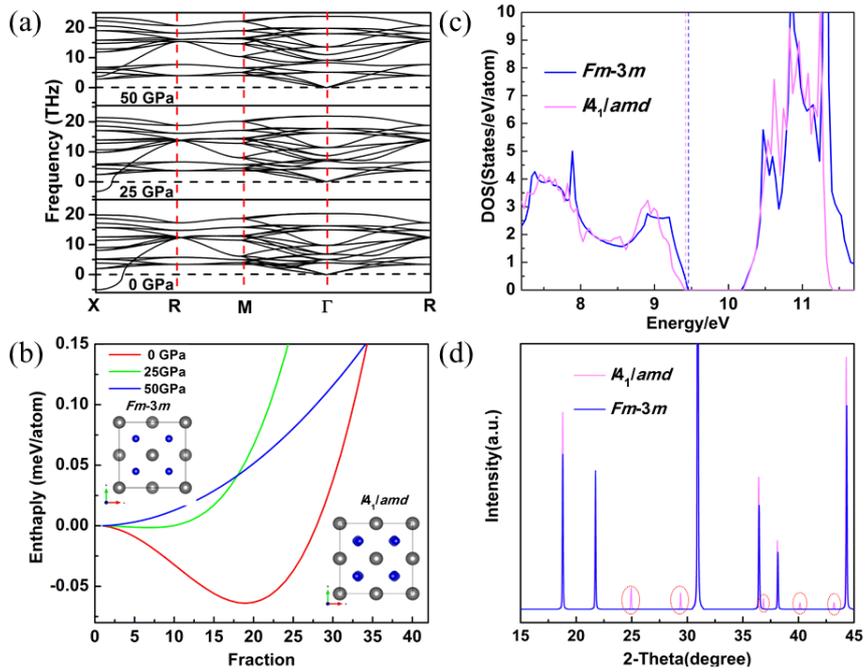

Fig.6 (a) The phonon dispersion of UN$_2$ at 0 GPa, 25 GPa, and 50 GPa; (b) (c) density of states of UN$_2$ in $I4_1/acd$ and $Fm$-$3m$ phases. The corresponding Fermi energy in both cases is represented by dashed line. (d) The simulated XRD for UN$_2$ in $I4_1/acd$ and $Fm$-$3m$ phases. (a) The simulated XRD patterns for UN$_2$ in $I4_1/acd$ and $Fm$-$3m$ phases with a wavelength of 1.54056 Å at 0 GPa

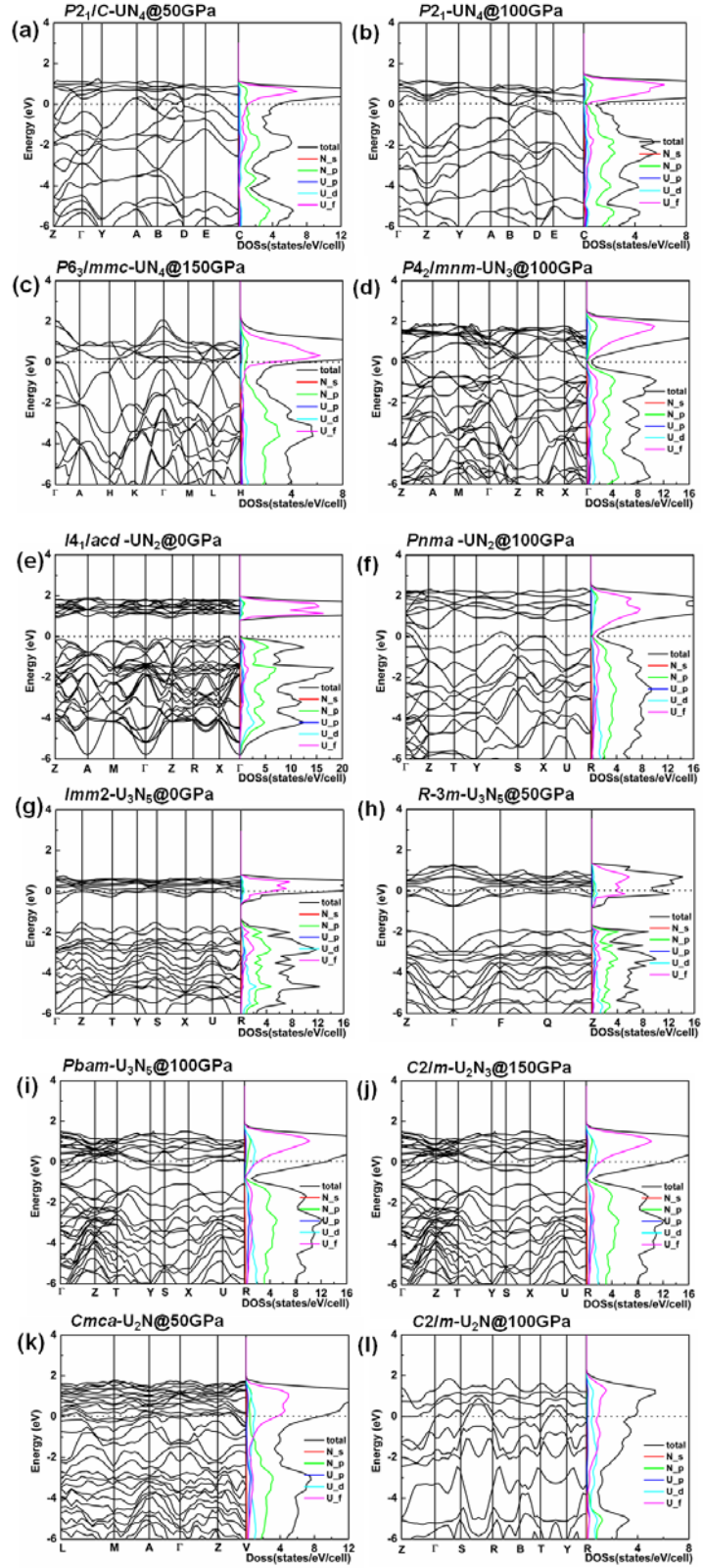

Fig.7. The band structures and the density of states (DOSs) in newly found $U_mN_n$ compounds at selected pressures.

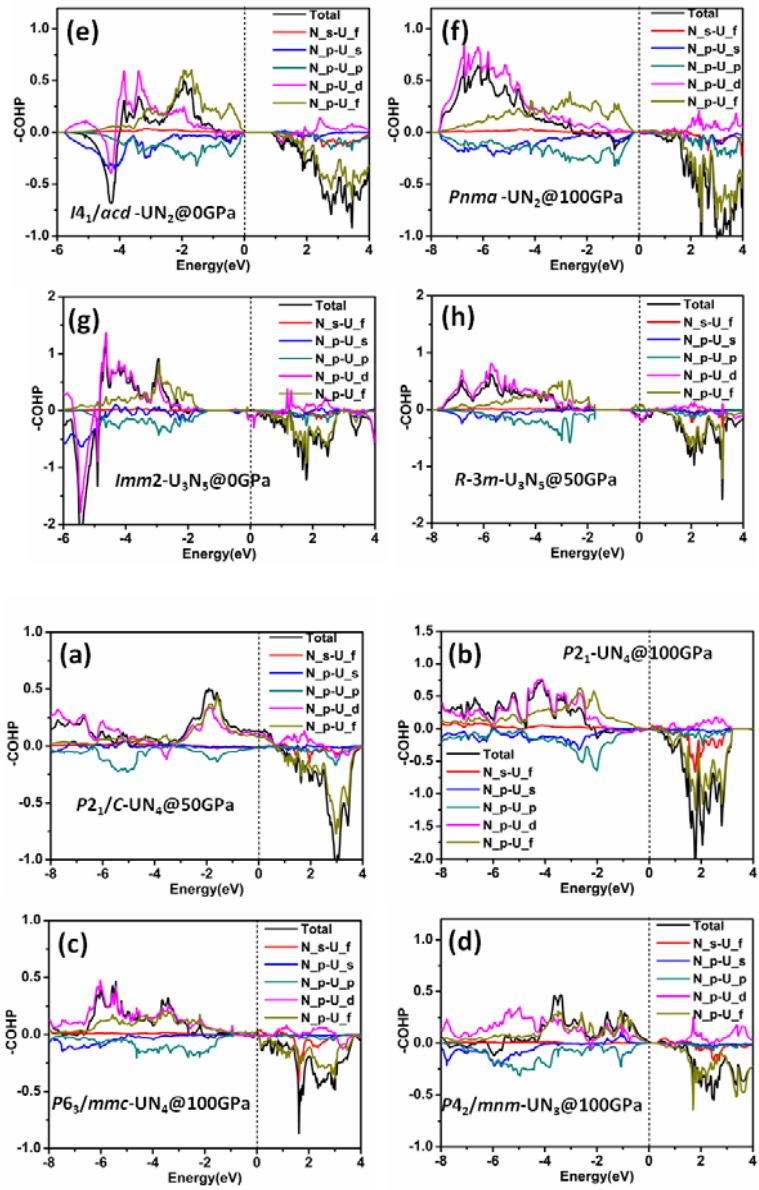

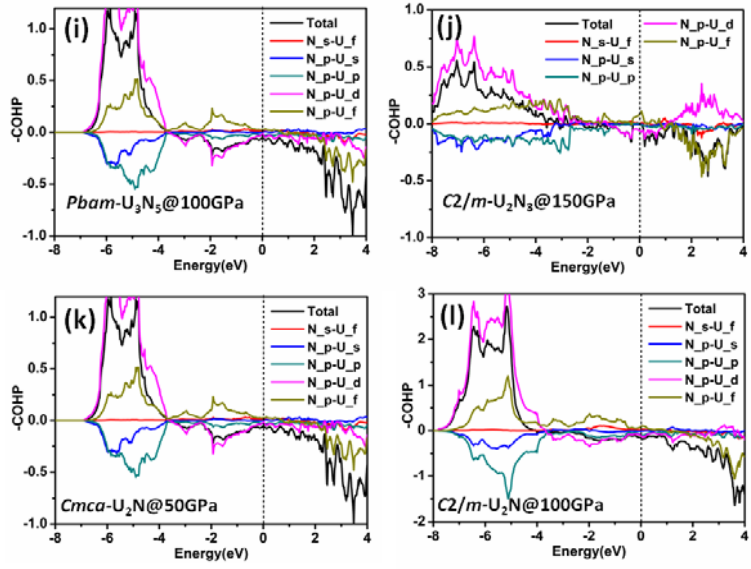

**Fig.8**. COHP functions for U-N interactions in newly-predicted UN compounds at selected pressure. (a) $UN_4$ at 50 GPa in the *$P2_1/C$*. (b) $UN_4$ at 100 GPa in the *$P2_1$*. (c) $UN_4$ at 150 GPa in the *$P6_3/mmc$*. (d) $UN_3$ at 100 GPa in the *$P4_2/mnm$*. (e) $UN_2$ at 0 GPa in the *$I4_1/acd$*. (f) $UN_2$ at 100 GPa in the *Pnm*a. (g) $U_3N_5$ at 0 GPa in the *Imm2*. (h) $U_3N_5$ at 50 GPa in the *R-3m*. (i) $U_3N_5$ at 100 GPa in the *Pbam*. (j) $U_2N_3$ at 150 GPa in the *C2/m*. (k) $U_2N$ at 50 GPa in the *Cmca*. (l) $U_2N$ at 100 GPa in the *C2/m*.

**Surporting Informations**

# Prediction of Stable Ground-State Uranium Nitrides at Ambient and High Pressures

Dawei Zhou[†], JiaHui Yu[‡], Chunying Pu*[,†], Yuling Song[†]

[†] *College of Physics and Electronic Engineering, Nanyang Normal University, Nanyang 473061, China*

[‡] *Department of Electronics and Electrical Engineering, Nanyang Institute of Technology, Nanyang 473004, China*

Corresponding authors: Email: puchunying@126.com

**Table S1.** The calculated lattice parameters (in Angstroms) and atomic positions for the newly found $U_mN_n$ compounds at selected pressure.

| | Space group | Pressure (GPa) | lattice parameters | Atom | Site | Wyckoff | positions | |
|---|---|---|---|---|---|---|---|---|
| $UN_4$ | $P2_1/C$ | 50 | $a=3.5588$<br>$b=5.8148$<br>$c=4.1823$<br>$\alpha=\gamma=90.0$<br>$\beta=73.555$ | N | 4e | 0.07819 | 0.74007 | 1.37383 |
| | | | | N | 4e | 0.74651 | 0.63975 | 1.44608 |
| | | | | U | 2d | 0.50000 | 0.50000 | 1.00000 |
| | $P2_1$ | 100 | $a=3.5282$<br>$b=5.7536$<br>$c=3.8258$<br>$\alpha=\gamma=90.0$<br>$\beta=73.8963$ | N | 2a | 0.56634 | 0.08952 | 0.68415 |
| | | | | N | 2a | 0.24355 | 0.97374 | 0.68722 |
| | | | | N | 2a | 0.23432 | 0.21207 | 0.20179 |
| | | | | N | 2a | 0.43782 | 0.59080 | 0.94651 |
| | | | | U | 2a | 0.01886 | 0.35118 | 0.74151 |
| | $P6_3/mmc$ | 150 | $a=b=4.4829$<br>$c=3.9077$<br>$\alpha=\beta=90.0$<br>$\gamma=120$ | N | 6h | 0.66332 | 0.83166 | 0.25000 |
| | | | | N | 2b | 0.0000 | 0.0000 | 0.75000 |
| | | | | U | 2d | 0.33333 | 0.66667 | 0.75000 |
| $UN_3$ | $P4_2/mnm$ | 100 | $a=b=5.5028$<br>$c=4.2292$<br>$\alpha=\beta=\gamma=90.0$ | N | 8i | 0.82898 | 0.65030 | 0.00000 |
| | | | | N | 4e | 0.50000 | 0.50000 | 0.33809 |
| | | | | U | 4g | 0.73681 | 0.26319 | 0.00000 |
| $UN_2$ | $I4_1/acd$ | 0 | $a=b=5.3052$<br>$c=10.6153$<br>$\alpha=\beta=\gamma=90.0$ | N | 32g | -0.27144 | -0.75000 | 0.37500 |
| | | | | U | 16c | 0.00000 | -0.50000 | 0.00000 |
| | $Pnma$ | 100 | $a=5.4377$<br>$b=3.2986$<br>$c=6.2014$<br>$\alpha=\beta=\gamma=90.0$ | N | 4c | -0.47849 | 0.25000 | 0.17152 |
| | | | | N | 4c | -0.13938 | 0.25000 | -0.93179 |
| | | | | U | 4c | 0.25309 | 0.25000 | 0.89205 |
| $U_3N_5$ | $Imm2$ | 0 | $a=11.3243$<br>$b=3.6915$<br>$c=5.3861$<br>$\alpha=\beta=\gamma=90.0$ | N | 4c | 0.32691 | 0.50000 | 0.35158 |
| | | | | N | 4c | 0.35590 | 0.50000 | -0.14895 |
| | | | | N | 2b | 0.50000 | 0.00000 | 0.31813 |
| | | | | U | 4c | 0.32558 | 0.00000 | 0.10714 |
| | | | | U | 2a | 0.00000 | 0.00000 | 0.07158 |
| | $R$-$3m$ | 50 | $a=b=3.6045$<br>$c=23.6668$<br>$\alpha=\beta=90.0$<br>$\gamma=120.0$ | N | 3b | 0.00000 | 0.00000 | -0.50000 |
| | | | | N | 6c | 0.00000 | 0.00000 | -0.90448 |
| | | | | N | 6c | 0.00000 | 0.00000 | -0.69746 |
| | | | | U | 6c | 0.00000 | 0.00000 | -0.78663 |
| | | | | U | 3a | 0.00000 | 0.00000 | 0.00000 |
| | *Pbam* | 100 | *a*= 4.9442<br>*b*=8.9632<br>*c*= 3.4325<br>α=β=γ=90.0 | N | 4g | 0.14425 | 0.27699 | 0.00000 |
| | | | | N | 4h | 0.65377 | 0.90158 | 0.50000 |
| | | | | N | 2a | 0.50000 | 0.50000 | 0.00000 |
| | | | | U | 2c | 0.50000 | 0.00000 | 0.00000 |
| | | | | U | 4h | 0.93259 | 0.16092 | 0.50000 |
| U$_2$N$_3$ | *C2/m* | 150 | *a*=11.1919<br>*b*=3.5719<br>*c*=4.5298<br>α=γ=90.0<br>β=104.9442 | N | 4i | 0.11526 | -1.00000 | 0.38147 |
| | | | | N | 2b | 0.00000 | -0.50000 | 1.00000 |
| | | | | N | 4i | -0.21071 | -0.50000 | 0.18646 |
| | | | | N | 2d | 0.00000 | -0.50000 | 0.50000 |
| | | | | U | 4i | 0.18237 | -0.50000 | 0.28923 |
| | | | | U | 4i | -0.07974 | 0.00000 | 0.18850 |
| U$_2$N | *Cmca* | 50 | *a*=3.2677<br>*b*=5.2438<br>*c*=9.5186<br>α=β=γ= 90.0 | N | 4b | 0.00000 | 0.50000 | 0.00000 |
| | | | | U | 8f | 0.00000 | 0.34791 | 0.64830 |
| | *C2/m* | 100 | *a*= 4.9534<br>*b*=3.1705<br>*c*=6.0804<br>α=γ=90.0<br>β=129.7890 | N | 2b | -0.50000 | 0.00000 | 0.00000 |
| | | | | U | 4i | -0.91810 | 0.00000 | -0.70573 |

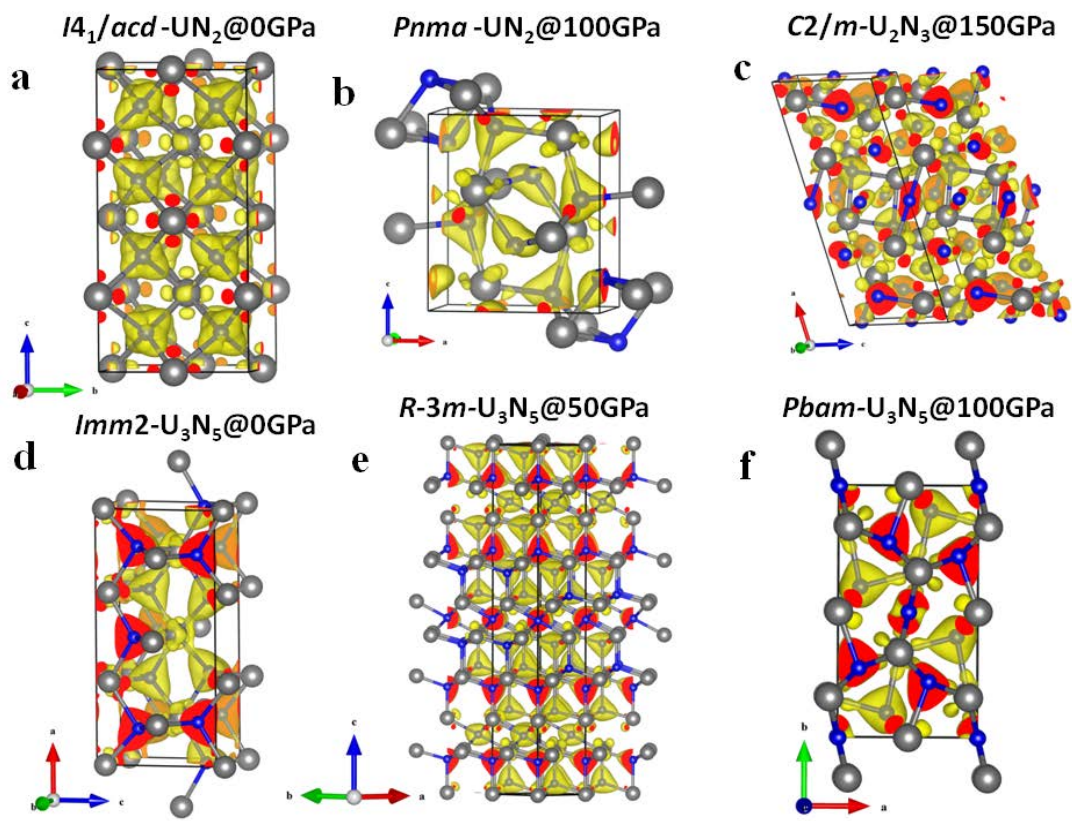

**Fig.S1.** Difference charge density maps for predicted U-N compunds at selected pressure. (a) $I4_1/acd$-UN$_2$ at 0 GPa. (b) $Pnma$-UN$_2$ at 100 GPa. (c) $Imm2$-U$_3$N$_5$ at 0 GPa. (d) $R$-$3m$-U$_3$N$_5$ at 50 GPa. (e) $Pbam$-U$_3$N$_5$ at 100 GPa. (f) $C2/m$-U$_2$N$_3$ at 150 GPa.

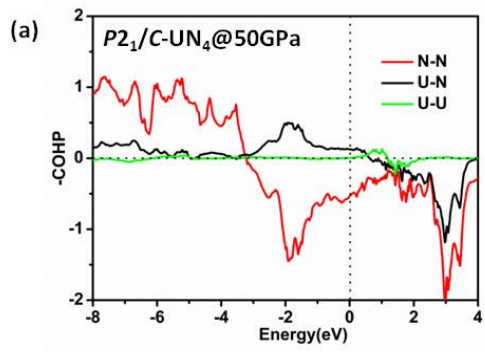
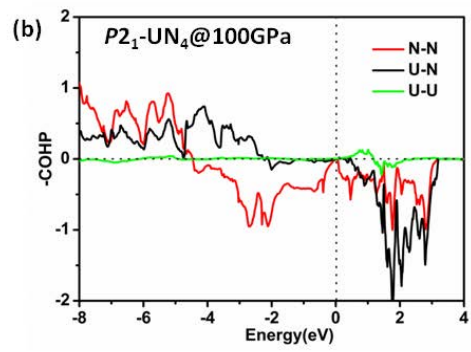
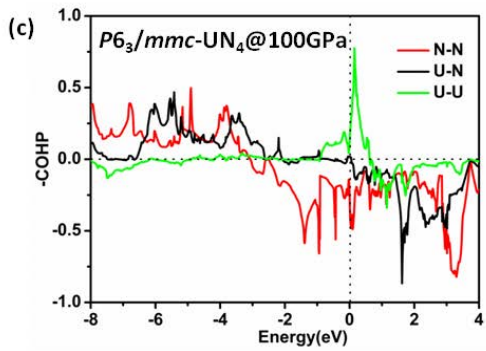
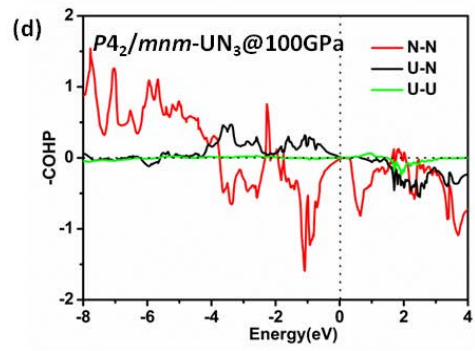
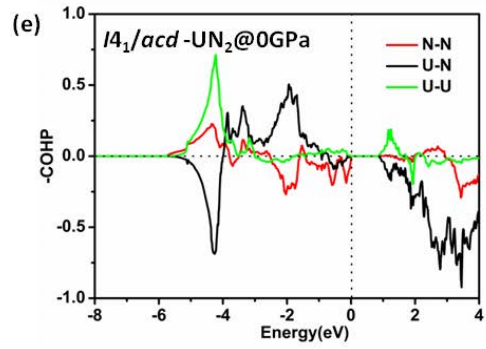
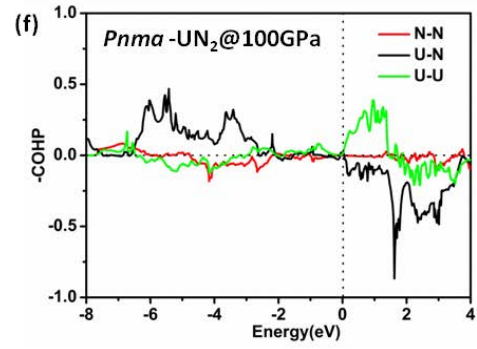
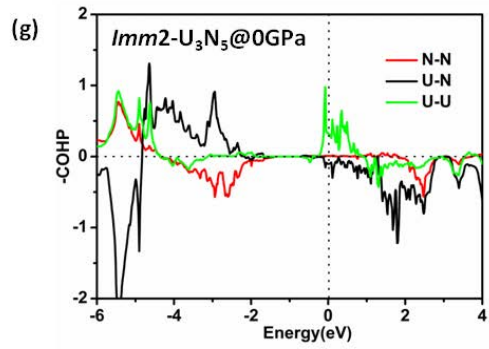
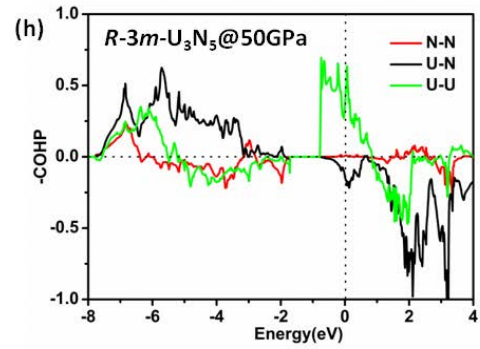

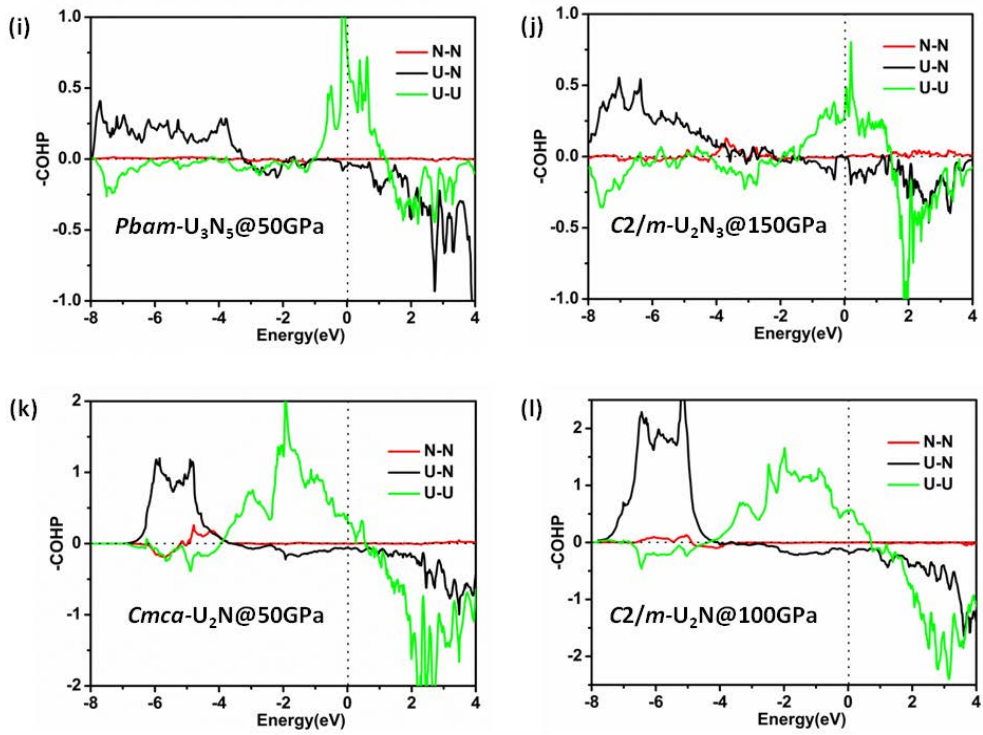

**Fig.S2.** COHP functions for N-N, U-N, and U-U interactions in newly predicted U-N compounds at selected pressure. (a) $P2_1/C$-$UN_4$ at 50 GPa. (b) $P2_1$-$UN_4$ at 100 GPa. (c) $P6_3/mmc$-$UN_4$ at 150 GPa. (d) $P4_2/mnm$-$UN_3$ at 100 GPa. (e) $I4_1/acd$-$UN_2$ at 0 GPa. (f) $Pnma$-$UN_2$ at 100 GPa. (g) $Imm2$-$U_3N_5$ at 0 GPa. (h) $R$-$3m$-$U_3N_5$ at 50 GPa. (i) $Pbam$-$U_3N_5$ at 100 GPa. (j) $C2/m$-$U_2N_3$ at 150 GPa. (k) $Cmca$-$U_2N$ at 50 GPa. (l) $C2/m$-$U_2N$ at 100 GPa. The Fermi level is at zero energy.